\documentclass{elsart}
\begin{document}
\begin{frontmatter}

\title{The concept of Existence \\ and the derivation of a Quantum Force}
\vspace{.1in}
\author{{\sc C. Costa}}
\address{PUC-Rio, CP 38071,
Rio de Janeiro, RJ 22452-970, Brazil}
\vspace{.5in}
\begin{abstract}
We define a new dynamical variable, the relative existence $e$, in
terms of space and time. Taking it as a generalized positional
coordinate, we show that for conservative systems the canonically
conjugated momentum is identified as the classical force. Applying
Wilson-Sommerfeld-Bohr's quantum conditions for a conditionally
periodic motion, we derive an expression for the quantum force, $F
\,= \hbar k f_e$, where $k$ is the wave number and $f_e$ is the
characteristic frequency of the system. Applying Dirac's method to
the Poisson Brackets involving existence and force, we obtain the
uncertainty relation $\Delta e \, \Delta F \,\ge \hbar / 2$.
\\ The force quantization may have already been observed in stimulated
bichromatic optical force experiments, used to deflect, decelerate
and manipulate laser-cooled atomic beams.
\end{abstract}
\end{frontmatter}

\newpage
\section{Introduction}
Based on Planck-Einstein relation for the quantized energy of a
photon, $E = h \nu$, and the results of Rutheford experiments on
the scattering of alpha particles by matter, Niels Bohr was able
to propose the model for the hydrogen atom, defining discontinuous
stationary states, so that emission or absorption of radiation
corresponds to transitions between such quantized
states\cite{bohr:1913}. The discrete energy levels can be
calculated by means of a set of quantum conditions to which the
canonical variables $q_i$ and $p_i$ of classical mechanics were to
be subjected\cite{bohr:1918}. Even filled with limitations and
being a mix of classical and quantum concepts, Bohr's ``old
quantum theory'' opened up the way to the Quantum Mechanics.

In this paper we propose a new dynamical variable, related to
space and time, which we call {\it relative existence}. Regarding
existence as a canonical positional variable, we follow
wilson-Sommerfeld-Bohr's prescription, looking after its canonical
conjugate momentum and writing the corresponding quantum
conditions. We find that for conservative systems, like the simple
harmonic oscillator, the variable conjugate to existence is the
classical force, so that the quantum conditions lead to the
expression for a quantum force. Applying Dirac's method of
quantization\cite{dirac:1925}, it is possible to write the
existence-force uncertainty relation. Finally we discuss the
possibility that force quantization may have already been
manifested in the stimulated bichromatic optical
force\cite{voitsekhovich,nolle:1996,soding:1997,goepfert:1997},
measured as a deflection or deceleration of atomic beams subjected
to a field of two counterpropagating short laser pulses, in laser
cooling and atom manipulation experiments
\cite{metcalf:Rb,metcalf:He,metcalf:2003}.

\section{Dynamical variables: defining relative existence}
In classical mechanics, the one-dimensional acceleration $a$ is
defined as the derivative of velocity $v$ with respect to time
$t$, $a=dv/dt$, so that $v= v_{0} + \int a \,dt$. Similarly,
velocity is defined as the derivative of the position $x$ with
respect to time, $v=dx/dt$, so that $x= x_{0} + \int v \,dt$.

We now define a new variable, say $e$, enabling us to write the
position as the derivative of $e$ with respect to time,
\begin{eqnarray}
 x &=&\frac{de}{dt},  \mbox{\hspace{0.5cm}so that:\hspace{0.5cm}} \\
 e &=& e_{0} + \int x \; dt.  \label{eq:e_def}
\end{eqnarray}

This variable is related to a body occupying some place in space,
extended over the time, and we suggest to name it ``relative
existence'', or simply {\it existence}. Even not moving, the body
has some degree of existence. As it moves or it accelerates, more
and more intense is the accumulation of existence, as for the mass
in special relativity. The absolute value of the variation of
existence shall be called {\it experience} ($X_p \equiv |\,e -
e_0\,| = |\,\Delta e\,|\,$), in a fashion similar to the
displacement, which is regarded as the absolute value of the
variation of position, or to the impulse, the variation of
momentum. We shall therefore speak of particles being created at
particular space-time coordinates and evolving their existences,
accumulating experience more or less rapidly, depending on their
behavior.

From definition (\ref{eq:e_def}) we obtain the relativistic
equation for the existence, in terms of Lorentz Transformations.
Let the object denoted by 0 move with respect to the reference
system 1, with velocity $v_{01}$ at the position $x_{01}$, then
for a second system 2, moving with relative velocity $u_{21}$ with
respect to 1, we have:
\begin{eqnarray}
v_{02} & = & \gamma^0 \; \left( 1 +
\frac{v_{01}u_{12}}{c^2}\right)^{-1} \; \left(v_{01} +
u_{12}\right), \\
x_{02} & = & \gamma^1 \; \left( 1 +
\frac{v_{01}u_{12}}{c^2}\right)^{0} \; \left(x_{01} +
(c t_{01})\frac{u_{12}}{c}\right), \\
c t_{02} & = & \gamma^1 \; \left( 1 +
\frac{v_{01}u_{12}}{c^2}\right)^{0} \; \left( c t_{01} +
(x_{01}) \frac{u_{12}}{c}\right), \\
e_{02} & = & \gamma^2 \; \left( 1 +
\frac{v_{01}u_{12}}{c^2}\right)^{1} \; \left(e_{01} +
e_{12}\right);
\end{eqnarray}

\noindent where $\gamma =(1-u^2/c^2)^{-1/2}$ is the Lorentz
factor, and the equations are written in a way to stress the
relationship between the powers of different factors.

\section{Canonically conjugate variable: force}
Let us assume that the existence may be regarded
as a generalized coordinate $q$:
\begin{equation} q=e ,\end{equation}

\noindent then, the generalized velocity is
\begin{equation}
\dot{q} \; = \; \frac{de}{dt} \; \equiv \; x.
\end{equation}

In terms of the Lagrangian $\mathcal{L}$ of the system, the
generalized momentum $p_e$, conjugate to $q=e$, is
\begin{equation}
p_e \; = \; \frac{\partial \mathcal{L}}{\partial \dot{q}} \;= \;
\frac{\partial \mathcal{L}}{\partial x} . \label{eq:pe_dldx}
\end{equation}

For conservative systems the Lagrangian is given by $\mathcal{L}$
$= T - V(x)$, where $T$ is the kinetic energy, with no explicit
dependence on the position $x$, and $V(x)$ is the potential
energy. In this case,
\begin{equation}
p_e \; = \; \frac{\partial \mathcal{L}}{\partial x} \;= \; -
\frac{\partial V(x)}{\partial x} \equiv F , \label{eq:pe_f}
\end{equation}

\noindent meaning that we may identify the canonical conjugate to
the existence $e$ as the classical force $F$.

The Simple Harmonic Motion (SHM) easily illustrates equation
(\ref{eq:pe_f}). Hook's law for a spring acting upon a particle of
mass $m$ gives:
\begin{equation}
F \; = \; -kx \; = -(m\omega^2)x , \label{eq:f_kx}
\end{equation}

\noindent where $k=m\omega^2$ is the effective spring constant and
$\omega$ is the harmonic oscillator angular frequency. Newton's
Law then implies the equation of motion given by $\ddot{x} = -
\omega^2 x$, with the known solution of the form $x(t) = x_o
\cos(wt +\phi)$. Assuming $e_o =0$ at $t=0$, the time evolution of
the existence is
\begin{equation}
e(t) \; = \; -\frac{1}{\omega^2} \; \dot{x}(t). \label{eq:e_dotx}
\end{equation}

Since $V(x) = \frac{1}{2}\, k x^2 = \frac{1}{2}\, m \omega^2 x^2$,
we can write the Lagrangean for the SHM,
\begin{equation}
{\mathcal{L}} (x,\dot{x}) \;=\; T-V \;=\; \frac{1}{2} m \dot{x}^2
-\frac{1}{2} m \omega^2 x^2 ,
\end{equation}

\noindent in terms of the generalized coordinates $q=e$ and
$\dot{q}=x$:
\begin{equation}
{\mathcal{L}} (e,x) \;=\; \frac{1}{2} m \omega^4 e^2 -\frac{1}{2}
m \omega^2 x^2 .
\end{equation}

The canonical conjugate momentum is therefore
\begin{equation}
p_e \;\equiv\; \frac{\partial \mathcal{L}}{\partial \dot{q}} \;=\;
\frac{\partial \mathcal{L}}{\partial x} \;=\; -m \omega^2 x ,
\label{eq:pe}
\end{equation}

\noindent which is readily recognized as the classical force,
eq.(\ref{eq:f_kx}).

Moreover, the generalized force $Q_e$ is, here,
\begin{equation}
Q_e \;\equiv\; \frac{\partial \mathcal{L}}{\partial q} \;=\;
\frac{\partial \mathcal{L}}{\partial e} \;=\; m \omega^4 e ,
\label{eq:qe}
\end{equation}

\noindent so that the Euler-Lagrange equation,
\begin{equation}
\frac{d}{dt}\left( \frac{\partial \mathcal{L}}{\partial
\dot{q}}\right) \;=\; \frac{\partial \mathcal{L}}{\partial q},
\mbox{\hspace{0.5cm}or:\hspace{0.5cm}}
\frac{d}{dt}\left(p_e\right) \;=\; Q_e,
\end{equation}

\noindent is verified for the existence-force conjugates, in view
of equations (\ref{eq:pe}), (\ref{eq:qe}) and (\ref{eq:e_dotx}):
\begin{equation}
\frac{d}{dt}\left(p_e\right) \;=\; -m \omega^2 \dot{x} \;=\; -m
\omega^2 \left(-\omega^2  e\right) \;=\; m \omega^4 \,e \;=\; Q_e.
\end{equation}

Similarly, it is straightforward to verify that the Hamilton
Canonical Equations also hold.

\section{Bohr's quantum conditions: deriving the quantum force}
For any physical system in which the coordinates are periodic
function of time, there exist a quantum condition for each
coordinate. These are the Wilson-Sommerfeld-Bohr quantum
conditions, constraining the classical action integral $I_k$
according to\cite{bohr:1918}:
\begin{equation}
I_k \;=\; \oint p_k \, dq_k \;=\; n_k \, h, \label{eq:i_k}
\end{equation}

\noindent where the quantum number $n_k$ is an integer, $p_k$ is
the momentum associated to the generalized coordinate $q_k$, and
each integral is taken over a full period of this coordinate.

Applied to the circular motion with a central field force,
corresponding to the angle-angular momentum conjugate variables
$(\theta-L)$, Bohr obtained the angular momentum quantization for
the Hydrogen atom as
\begin{equation}
L \,=\, n \, \frac{h}{2\pi} \,=\, n\,\hbar, \label{eq:l_theta}
\end{equation}

\noindent and derived thereafter the quantization of atomic orbits
and the correct energies for the line-spectra. Regarding the
time-energy pair $(t-E)$ as conjugates, eq.(\ref{eq:i_k}) leads
directly to Planck-Einstein photon energy quantization,
\begin{equation}
E \,=\, h \nu. \label{eq:e_hnu}
\end{equation}

\noindent For the position-momentum conjugates $(x-p)$, we obtain
the matter wave relation of De Broglie,
\begin{equation}
p \,=\, \frac{h}{\lambda} \,=\, \hbar k. \label{eq:p_hlambda}
\end{equation}

Considering eq.(\ref{eq:i_k}) applied to the existence-force
conjugates $(e-F)$, we have
\begin{equation}
\oint F \, de \;=\; n \, h. \label{eq:f_he}
\end{equation}

Inspired by equations (\ref{eq:l_theta}), (\ref{eq:e_hnu}) and
(\ref{eq:p_hlambda}), and following an {\it ansatz} similar to De
Broglie's, we may write
\begin{equation}
F \, \oint de \;=\; F \, \oint x \,dt \;=\; F \,\lambda\,\tau_e ,
\label{eq:f_de}
\end{equation}

\noindent where $\lambda$ is the associated wavelength (so that
$k= 2\pi /\lambda$) and $\tau_e=1/f_e$ is the characteristic time
of the system, with characteristic frequency $f_e$. We therefore
write:
\begin{equation}
\tilde{F} \; \equiv \; n\frac{h}{\lambda\,\tau_e} \;=\; n\,\hbar
\, k \,f_e . \label{eq:quantumF}
\end{equation}

This is to be interpreted as an expression for the force
quantization, so that the unit quantum force is $\tilde{F} = \hbar
k f_e$, where we used the tilde in $\tilde{F}$ to remark its
quantized nature and to distinguish from the classical counterpart
$F$.

\section{Dirac's quantum conditions: uncertainty relation}
Poisson Brackets (PB) of two state variables $a(q,p)$ and $b(q,p)$
are defined as:
\begin{equation}
\left\{ a(q,p) \, , \, b(q,p) \right\} \equiv \left(
\frac{\partial a}{\partial q} \frac{\partial b}{\partial p} -
\frac{\partial a}{\partial p} \frac{\partial b}{\partial q}
\right) .
\end{equation}

The PB between the canonical coordinates and momenta, $q$ and $p$,
is simply $\{q,p\} = 1$.

In Dirac's point of view\cite{dirac:1925}, the PB's are the
classical counterpart of the commutation rule of quantum
operators, establishing a rule for the classical-quantum
connection. Dirac defined the general quantum conditions, or
commutation relations, for any two quantum quantities $a$ and $b$:
\begin{equation}
\left[ a \, , \, b \right] \,=\, ab - ba \nonumber \,\equiv\, \pm
i \, \hbar \, \left\{ a \, , \, b \right\} ,
\end{equation}

\noindent which for the canonical conjugates reduces to Dirac's
fundamental quantum condition, $\left[q \, , \, p \right] \,=\,
\pm i \, \hbar$. It is possible to show\cite{cohen:qm} that
whenever the fundamental condition holds, one can write the
corresponding Heisenberg uncertainty relation $\Delta q \, \Delta
p \, \geq \, \hbar/2$.

Applied to the existence-force canonical conjugates $(e-F)$, we
obtain equivalent relations:
\begin{eqnarray}
\left\{ e \, , \, F \right\} &=&1 , \label{eq:ef_poissonb}\\
\left[ e \, , \, F \right] \; &=& \pm i \, \hbar , \label{eq:ef_commut}\\
\Delta e \, \Delta F &\geq& \frac{\hbar}{2}.
\label{eq:ef_uncertainty}
\end{eqnarray}

Take for example the SHM described before.
Eq.(\ref{eq:ef_poissonb}) is readily satisfied, as for any
canonical conjugate pair.  To verify eq.(\ref{eq:ef_commut}), we
just write the relations obtained classically:
\begin{eqnarray}
q \;=\; e &=& \frac{\dot{x}}{-\omega^2} \;=\; \frac{p}{-m\omega^2}
\nonumber \\
p_e \;=\; F &=& m\ddot{x} \;=\; -m\omega^2 x ,
\end{eqnarray}

\noindent and notice that in this particular case, the quantum
mechanical counterparts may give:
\begin{eqnarray}
\left[ e \, , \, F \right] &=& \left[ \frac{p}{-m\omega^2} \, , \,
-m\omega^2 x  \right] \nonumber \\ &=& \left[ p \, , \, x \right]
\, = \, - \, i \, \hbar .
\end{eqnarray}

From similar reasoning we conclude that, for the SHM, we may also
have $\Delta e \, \Delta F \,=\, \Delta p \, \Delta x \;\geq
\hbar/2$ , and the limiting condition (\ref{eq:ef_uncertainty}) is
contemplated.

\section{Experimental evidence: bichromatic optical force}
We shall look for an experimental manifestation of quantum force
effects. The quantum force may act under resonant conditions; and
interaction of photons with atoms may be the primary source to
produce such conditions. Ultra-short laser pulses may deal with
small amounts of momentum transfer; and laser-cooled atoms may be
suitable to investigate such exchanges minimizing non-quantum
effects (like Doppler broadening). Considering all these, we are
led to a series of experiments involving the stimulated
bichromatic optical force on atoms
\cite{voitsekhovich,nolle:1996,soding:1997,goepfert:1997}, used in
laser cooling and atom manipulation experiments
\cite{metcalf:Rb,metcalf:He,metcalf:2003}.

Excited atoms return to their ground states either by spontaneous
decay or stimulated emission. Spontaneous decay results in the
dissipative radiative force, whose magnitude saturates at
\begin{equation}
F_{rad}\;=\;\hbar \,k \,\Gamma /2, \label{eq:f_rad}
\end{equation}

\noindent where $\Gamma = 1/\tau$ is the natural linewidth,
related to the excited state lifetime $\tau$. Doppler cooling of
two-level atoms has been achieved by an incoherent sequence of
absorption followed by spontaneous emission, and the forces in the
process are therefore limited to $F_{rad}$. Laser cooling in
multilevel atoms is achieved by the rapid coherent sequence of
absorption followed by stimulated emission, through the use of
multiple beams of monochromatic light, taking advantage of the
dipole force working on the atoms. It can be shown that such
sub-Doppler cooling forces are similarly limited to a magnitude of
$\hbar k \Gamma /2$. There has been the theoretical
prediction\cite{kazantsev:1974} that a force in a field of two
counterdirected short laser pulses, each containing two
frequencies, could generate a force greater than the spontaneous
one. From 1994 to 1997 this force has been demonstrated
experimentally as deflection or deceleration of molecular
Na$_{2}$\cite{voitsekhovich}, atomic Na\cite{nolle:1996} and
atomic Cs beams\cite{soding:1997,goepfert:1997}. Later on,
experiments producing and measuring the so-called bichromatic
optical force on atomic Rb\cite{metcalf:Rb} and
He\cite{metcalf:He} were performed and the mechanism was studied
in more detail \cite{metcalf:2003}.

The stimulated optical force is interpreted in terms of photon
transfer induced by optical $\pi$-pulses\cite{soding:1997}: the
light field can be viewed as two counterpropagating trains of
resonant beat pulses. Each light beam contains the two frequencies
$\omega \pm\delta$, detuned from the atomic resonance $\omega$ by
$\pm\delta$ (hence the difference is $2\delta$). The produced beat
frequency is $1/T=\delta/\pi$, with the amplitude-modulation
period $T\ll\tau$. Their equal intensities $I$ are chosen so that
the so-called $\pi$-pulse condition is fulfilled: the Rabi
frequency associated with a single monochromatic traveling wave is
set to $\Omega_R=\frac{\pi}{4} \delta$, where
$\Omega_R=\Gamma\sqrt{I/2I_s}$ , with saturation frequency
$I_s=\pi hc/(3\lambda^3\tau)$. Under such condition, the
probability of absorption (or of stimulating emission from the
excited atom) is unity. The effect of each beat pulse is to invert
the atomic population. The pulse trains can thus force an atom to
cycle between its lower and upper state in such a way that it
repeatedly absorbs photons from one wave and emits them into the
other one. A total sharp momentum of $2\hbar k$ is thus
transferred by one set of double $\pi$-pulses to every atom and
the transfer rate is given by the beat frequency, so that in
principle the corresponding optical force is $2 \hbar k/ T=2\hbar
k\delta/\pi$, much larger than $F_{rad}$. In practice, the
appropriate optical force is achieved for the laser intensity
under the condition $\Omega_R\simeq\delta$, rather than
$\Omega_R=\pi\delta/4$, taking into account geometrical
limitations that determine the laser beam size.

We now consider the stimulated light force experience in the
context of the quantum force derived in eq.(\ref{eq:quantumF}).
For a two-level atomic system $\lambda$ is just the atomic
transition wavelength. Under the $\pi$-pulse condition, the system
cycles between these two states, until it eventually decays. The
characteristic ``existence-time'' $\tau_e$ may not be just the
upper state lifetime $\tau$, because atoms are either in excited
or in ground state, and are being subjected to intense scattering
light field. Let the characteristic existence frequency
$f_e=1/\tau_e$ be identified with $\Gamma/2$, the maximum
scattering rate for a saturated optical transition with natural
linewidth $\Gamma = 1/\tau$. We are therefore led to write the
expression for the quantum force as
\begin{equation}
\tilde{F}\;\equiv\; \hbar \,k \,\Gamma /2. \label{eq:biquantumF}
\end{equation}

In spite of the equivalence between equations (\ref{eq:f_rad}) and
(\ref{eq:biquantumF}), the first equation sets a fundamental limit
over continuous radiative forces, while the second one sets the
quantum unit of interaction exchange.

Stimulated optical force experiments may have already seen the
manifestation of such force quantization. The early work from
N\"olle {\it et al.}\cite{nolle:1996} displays the spatial atom
distribution measured for the deflection of a thermal Na atomic
beam. First with a single pulse train, the fitted deflection force
for is given by $F=(0.8 \pm 0.1) \tilde{F}$, for most of the laser
intensity range investigated, respecting the limit given by
eq.(\ref{eq:f_rad}).  Switching a reflection mirror on, so that
the $\pi$-pulse condition is fulfilled, the deflection force
yields $F=(1.9 \pm 0.2) \tilde{F}$ and $F=(3.2 \pm 0.2)
\tilde{F}$, for two different laser intensities, in accordance to
eq.(\ref{eq:biquantumF}).

In both articles of Ref.\cite{metcalf:Rb}, the deflection of a Rb
atomic beam is measured and calculated for different setup
parameters (the relative phase of the pulses of the
counter-propagating light beams and the laser intensity). The
corresponding bichromatic optical force is then displayed as a
function of the atom´s velocity $v$. The optical force magnitude
is given in units of $\hbar k \Gamma$ (the authors use the
notation $\gamma$ instead of $\Gamma$) and we must multiply their
values by a factor two in order to get them in units of
$\tilde{F}=\hbar \,k \,\Gamma /2$. Preliminary assessment suggests
that the peak values of measured force are indeed about 1, 2, 3
and 4 units of $\tilde{F}$, indicating the manifestation of the
quantum force, in the domain of stimulated bichromatic optical
force experiments. In addition, these curves clearly present peaks
for different values of the velocity, although sometimes they do
overlap and become unresolved. We are tempted to assign a
resonance pattern to those curves, meaning that the quantization
of the force is correlated to discrete values of atomic beam
velocity - in a fashion similar to the quantization of Bohr´s
orbits due to angular momentum quantization. A numerical
correlation of the force quantization leading to velocity
quantization, for this particular system, is still to be
determined. One possibility is raised by the analysis of the
acceleration power of the light field, in these standing wave
field schemes, obtained by the stimulated conversion of photons
from one frequency component into the other. According to
S\"{o}ding {\it et al.}\cite{soding:1997}, the acceleration power
is limited by
\begin{equation}
F \, v \;=\; \hbar\Omega_R \,\Gamma /2. \label{eq:F_v}
\end{equation}

\noindent The quantum force requirement yields $\tilde{F}v = n\, v
\hbar k \,\Gamma /2$, which together with eq.(\ref{eq:F_v}), in
the saturated laser field domain, suggests a condition like
\begin{equation}
v\;=\; \frac{1}{n}\frac{\Omega_R}{k}. \label{eq:quantum_v}
\end{equation}

\noindent Such correlation shall yet be further investigated, and
confronted to the experimental data.

%
\section{Conclusion}
We have proposed the definition of a new dynamical variable, the
existence $e$, entangling space and time, with the interpretation
that any particle occupying certain position in space, as time
goes by, exists, accumulating experience. Besides any kinematical
consideration, interesting consequences arise when we assume the
existence as a generalized canonical coordinate, obtaining force
as the associated canonical momentum. Wilson-Sommerfeld-Bohr's
quantum conditions enabled us to derive the expression for a
quantum force, and Dirac's quantum prescription imposed an
Heisenberg-like limitation to it. A few results from the simple
harmonic oscillator are used to illustrate the existence-force
relationship.

We argue that the recent measurement of stimulated bichromatic
optical forces, observed as deflection of laser cooled atomic
beams subjected to short counter-propagating $\pi$-pulses of
laser, having magnitude several larger then the maximum limit
obtainable by radiative forces, may contain indication of the
force quantization, in terms of multiples of $\tilde{F} =\hbar \,
k \, \Gamma/2$. It may be not surprising that the force in these
experiments is actually quantized, since the momentum transfer
between the laser field and the atomic system is performed
coherently in a stimulated fashion.

Future experiments in the field of bichromatic optical forces
deserves to be carefully proposed with the purpose to investigate
these conjectures on quantum force in closer detail. A possible
correlation between quantized force and discrete values of atomic
velocity resonant with the optical field momemtum transfer is also
to be verified.

%
\section*{Acknowledgements}
The author wishes to thank C. Salles for support and
encouragement.



\end{document}